\documentclass[aps,pre,preprint]{revtex4}

\usepackage{amssymb}
\usepackage{epsfig}
\usepackage{graphicx}

\begin{document}


\title{Quiescence: a mechanism for escaping the effects of drug on cell populations}


\author{Tom\'as Alarc\'on}
\email[]{alarcon@bcamath.org}
\affiliation{Basque Centre for Applied Mathematics, Bizkaia Technology Park, 48160 Derio, Spain} 

\author{Henrik Jeldtoft Jensen}
\email[]{h.jensen@imperial.ac.uk}
\affiliation{Institute for Mathematical Sciences,   Imperial College London, 53 Princes Gate, London SW7 2PG, UK \& 
Department of Mathematics, Imperial College London, South Kensington Campus, London Sw7 2AZ, UK}


\date{\today}

\begin{abstract}  
We point out that a simple and generic strategy to lower the risk for extinction consists in
the developing a dormant stage in which the organism is unable to multiply but may die. The
dormant organism is protected against the poisonous environment. The result is to increase the
survival probability of the entire population by introducing a type of zero reproductive fitness.
This is possible, because the reservoir of dormant individuals act as a buffer that can cushion fatal
fluctuations in the number of births and deaths which without the dormant population would have
driven the entire population to extinction.
\end{abstract}


\maketitle


\begin{figure*}
\includegraphics[scale=0.5]{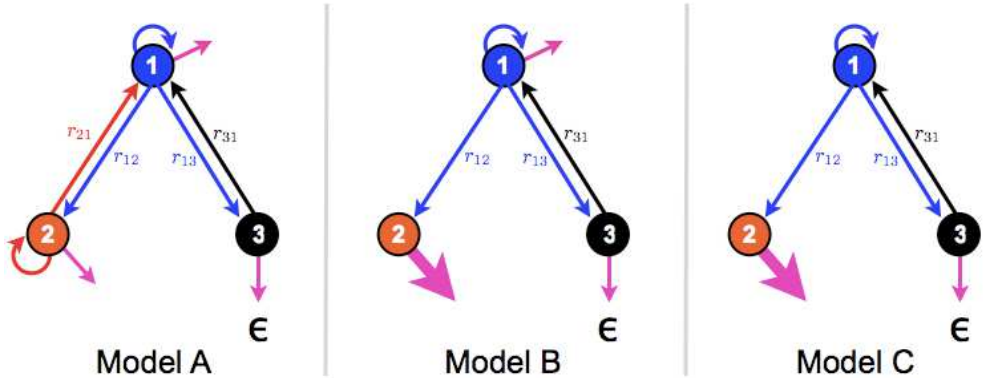}
\caption{Schematic representation of the population dynamics considered in this paper. The arrows between the different types of individuals represent the possible population flows between them. The arrows pointing to the types themselves represent proliferation. The other arrows stand for death of each type of individuals. The thickness of the arrows stand for the realtive magnitude of each of the flows or processes.\label{DiagA}}
\end{figure*}

\emph{Introduction.} Drug resistance is a very serious problem not least for chemotheaputic treatment of cancer and it is very important to understand, as far as possible, ways to countermeasure drug resistance in a number of biomedical contexts .  In order to unravel the possible mechanisms responsible for this phenomena Iwasa and coworkers developed a prominent theoretical model of the evolutionary dynamics of escape\cite{iwasa2003,iwasa2004}. They base their approach on the assumption that $n$ point mutations in some crucial parts of the genome are necessary for escape. They further assumed that the different mutants can be described by binary strings (with entries $+1$ or $-1$) of length $n$. There are $2^n-1$ such mutants. It is assumed that treatment reduces the proliferation ratios of sensitive mutants: $R<1$ whereas resistant mutants are such that $R>1$. The corresponding evolutionary dynamics is modelled in terms of Galton-Watson multitype branching process (GWMBP) \cite{kimmel2002} where at each generation each individual of each type has a given (in general, mutant-dependent) probability of mutating and producing offspring belonging to a different type. The problem is to calculate the probability that a resistant mutant is reached within a population of size $N$. The model proposed by Iwasa et al. has been analysed in more detail in \cite{serra2006,serra2007}. Whilst this approach is both biologically and mathematically sound, we 
believe the efficiency with which evolutionary escape allows organisms to avoid extinction, when attacked  by lethal drug, can be understood in a more simple way as a consequence of a dormant phase. To assess the generic importance of dormancy as an escape strategy we develop a very simple model framework, in which we only focus on the most essential aspects of population and evolution dynamics.

Here we show that the existence of a dormant phase can be crucial for a population to escape extinction. We consider the following simplified scenario. We study the survival probability of a population of organisms that can exist in the form of three different types, see the diagram \ref{DiagA}. The types differ in their response to the presence of a drug. Type (1) and type (2) have similar reproduction and death rates when no drug is present.  However, the drug is supposed to be lethal to type (2) but neutral to type (1). We are interested in how the existence of a dormant mode, type (3), effects the survival probability of the entire population.  The dormant type cannot reproduce, nor is it susceptible to the drug. However type (3) can die and it can undergo a transformation back to type (1). To understand the effect of the dormant type we consider different realisations (indicated in the diagram \ref{DiagA}) of the possible flow between the three different types.  

Our model is relevant to a number of biological cases, in particular to entities such as cancer cells, which  have been observed to evolve resistance to therapy: Treatments impose a selective pressure which eliminates the lesser fit strands of the corresponding populations but also drives an evolutionary process whereby better adapted individuals, i.e. individuals immune to the effects of the corresponding drug, eventually take over. Further rationale for our model is provided by the response of tumour cells to hypoxia (i.e. oxygen starvation). It is a well-known fact that hypoxia induces arrest of the cell-cycle \cite{gardner2001,alarcon2004}. This means that the rate at which cells replicate under hypoxia is drastically reduced, as hypoxia down-regulates the activity of the pathway regulating the progression through the cell-cycle \cite{gardner2001}. Moreover, another well-understood fact about cell survival/death regulation is that there exists cross-talk between the pathways regulating cell death and cell division \cite{nishioka1994,padmanabhan1999,alberts2002}. This means that, in normal circumstances, the down-regulation of the cell cycle machinery implies the down-regulation of the apoptotic (programmed cell death) machinery. It can be argued that such cross-talk is disrupted in cancer cells, and that cancer cells are such that cell death regulatory pathways are down-regulated. Therefore, under hypoxic conditions, cancer cells undergo a drastic reduction of their division and death rates. 

The study of the influence of hypoxic cancer cells fits rather well within the original remit in which the dynamics of evolutionary escape was put forward, as hypoxic cells are known to play a major role in the resistance to chemo- and radio-therapy in tumours \cite{bristow2008}. To study the influence of such a sub-population on the global dynamics of the total population, we model it as a dormant or quiescent population with neglegible proliferation rate and small death rate.

%

Previous works have analysed some of the effects of quiescent cells in tumour growth (see for example \cite{alarcon2003,brikci2008}). All these previous studies hint to the role of quiescence cells in the dynamics of tumour growth and its role in resistance to tumour growth but the issue of the evolutionary dynamics involved is not directly addressed. 

We now turn to the study of the survival probability. We disentangle the interplay between drug susceptibility, type (2), and dormancy, type (3), by analysing three different versions of the population dynamics, all are depicted in the diagram \ref{DiagA} and formulated as the following three Models.\\

\noindent{\bf Model A --} This is the "normal" situation, where no drug is present. Type (1) and type (2) are essentially equivalent, except that when type (1) reproduces it may undergo a "mutation" and end up as type (3). When type (2) reproduce it may mutate and become type (1).\\

\vspace{.2cm}
\noindent{\bf Model B --} This is the situation in the presence of the drug. The death rate of type (2) is now significantly bigger than the death rates of type (1) and type (3). Moreover, the drug makes type (2) unable to reproduce and therefore the flow from type (2) to type (1) is absent.

We will below find that the dormant stage increases the survival probability of the  population in the presence of the drug. To emphasis that the increased ability of the population to escape extinction is in fact caused by the presence of a dormant stage, we also consider an extreme version of the dynamics in which type (1) is unable to die. Namely,\\

\vspace{.2cm}
\noindent{\bf Model C --} This version of the the dynamics is equivalent to model B except that type (1) now is assumed not to be able to die directly, but has to flow through either type (2) or type (3) to do so. We then demonstrate that even in this extreme situation does the availability of the dormant stage enhance the population's chance for avoiding extinction.

We notice that the different types (1), (2) and (3) can be thought in epidemiological terms as susceptibles, infected and immune. The version of the dynamics defined as model B can be thought of as representing an age structured population. Type (3) is then juveniles, type (1) mature reproduction active individuals and type (2) are individuals in the post-reproductive stage.  

An economical and precise way to present the dynamics is in terms of the generator functions for the corresponding 
Galton-Watson multitype branching process, we include these generator functions in the tables in the appendix. Within the theory of multi-type branching process, the condition for eventual non-extinction is given in terms of the spectral radius, $\rho$, of the matrix ${\sf A}=(a_{ij})$, whose entries are the expected values of the number of offspring of type $j$ produced of an indvidual of type $i$. If $\rho>1$, there is a finite probability of $N(t\rightarrow\infty)>0$. The quantities $a_{ij}$ are calculate in terms of the corresponding generating functions (Table \ref{table:1}): $A_{ij}=\partial_jG_i\vert_{{\vec{x}={\bf 1}}}$. From these standard arguments of the theory of multi-type branching process, which carry on in a straightforward manner when size-dependence is taken into account \cite{jagers1999}, we can establish from Tables \ref{table:2} and \ref{table:1}, Appendix A, that the condition for asymptotic survival, that is for $P[N(t\rightarrow\infty)>0]>0$, is

\begin{equation}\label{eq:1}
2(1-r_{12})e^{-\mu N_{\infty}}>1,
\end{equation}

\noindent i.e. $r_{12}<1/2$. Otherwise, the probability of eventual extinction is 1. Consider now the presence of a third type (type 3) within the population, namely, quiescent individuals. These individuals are not allowed to proliferate but are resilient to the environment and can survive under hostile conditions, hence we assume that their death rate $\epsilon<<1$. In addition, quiescent cells are assumed to be able to revert back to type 1 at a given rate which depends on the availabity of resources. The issue we intend to analyse is whether the introduction of a quiescent sub-population helps to escape from the whole population being extinct. More precisely, the question we aim to address is: Assume  $r_{12}\geq 1/2$, is it possible for a population whose dynamics is described in the by Model B and Table \ref{table:1} to elude eventual extinction? Specifically this  scenario has been considered within the context of modelling of tumour growth, in particular in the response of cancer cells to hypoxia (low levels of oygen) \cite{alarcon2003}. Cancer cells appear to become quiescent in response to hypoxia, which is thought to give them an advatage in their competition with their normal counterparts as well as resistance to radio- and chemo-therpay \cite{brown2004}.

We now present the results of simulations of the survival probability for the three different scenarios represented by model A, B and C.

\begin{figure}
\includegraphics[scale=0.25]{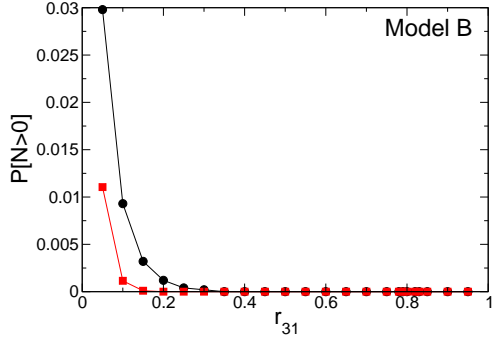}
\caption{Simulation results for $r_{12}=0.5$. Black circles correspond to $r_{13}=0.5$ and red squares to $r_{13}=0.25$. Other parameter values: $\mu=0.001$, $\epsilon=10^{-5}$, and $\alpha=1$.\label{fig:1}}
\end{figure}

\begin{figure}
\includegraphics[scale=0.25]{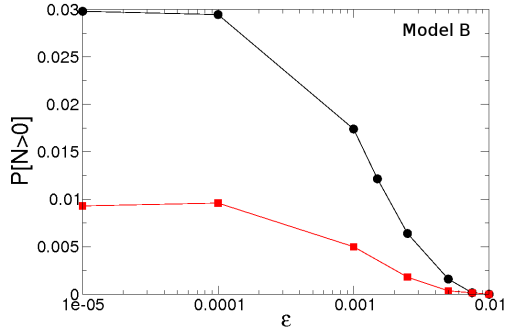}
\caption{Simulation results for $r_{13}=0.5$ and $r_{12}=0.5$, and green diamonds to $r_{12}=0.5$. Black circles correspond to $r_{31}=0.05$ and red squares $r_{31}=0.1$. Other parameter values: $\mu=0.001$, and $\alpha=1$.\label{fig:eps}}
\end{figure}

\emph{Simulation results.} Numerical simulations of the underlying multi-type branching process (see Appendix A for details) confirm that, in fact, the presence of a quiescent population as a mechanism for escape to harsh conditions is indeed feasible. Fig. \ref{fig:1} shows how the survival probability depends on the probability of an indvidual of type (1) to become quiescent, $r_{13}$. We can see that as $r_{13}$ decreases the threshold for survival in terms of $r_{31}$, i.e. the probability of a quiescent individual to revert to type (1), moves towards smaller values of the parameter $r_{31}$. This means that, following a decrease in the flux of individuals from the type (1) population to type (3), survival is only possible by reducing the inverse flux. This observation reveals that the mechanism by which quiescence helps escape is by acting as a reservoir where part of the population can be safely ``stored''. If the flux from type (3) back to type (1) is too big, it effectively increases the flux from type (1) to type (2), thus increasing lethality. This behaviour, however, is sensitive to the decay rate of type (3) individuals, $\epsilon$: as this parameter increases the survival probability decreases (see Fig. \ref{fig:eps}). This means that the death rate of quiescent individuals being small is an instrumental factor for quiescence-induced escape from harsh environments. 

\begin{figure}
\includegraphics[scale=0.25]{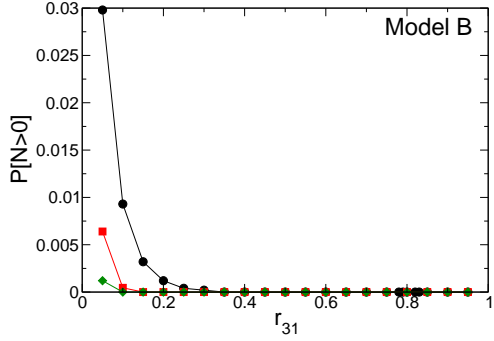}
\caption{Simulation results for $r_{13}=0.5$. Black circles correspond to $r_{12}=0.5$, red squares to $r_{12}=0.525$, and green diamonds to $r_{12}=0.55$. Other parameter values: $\mu=0.001$, $\epsilon=10^{-5}$, and $\alpha=1$.\label{fig:2}}
\end{figure}

Fig. \ref{fig:2} shows that as $r_{12}$ increases further into the regime where Eq. (\ref{eq:1}) predicts sure extinction of the (2)-type population, survival is only possible by decreasing $r_{31}$, i.e. by increasing the average time individuals spent in the quiescent state.
 
\begin{figure*}
\begin{center}
$\begin{array}{cc}
\mbox{(a)} & \mbox{(b)}\\
r_{13}=0 & r_{13}\neq 0\\
\includegraphics[scale=0.25]{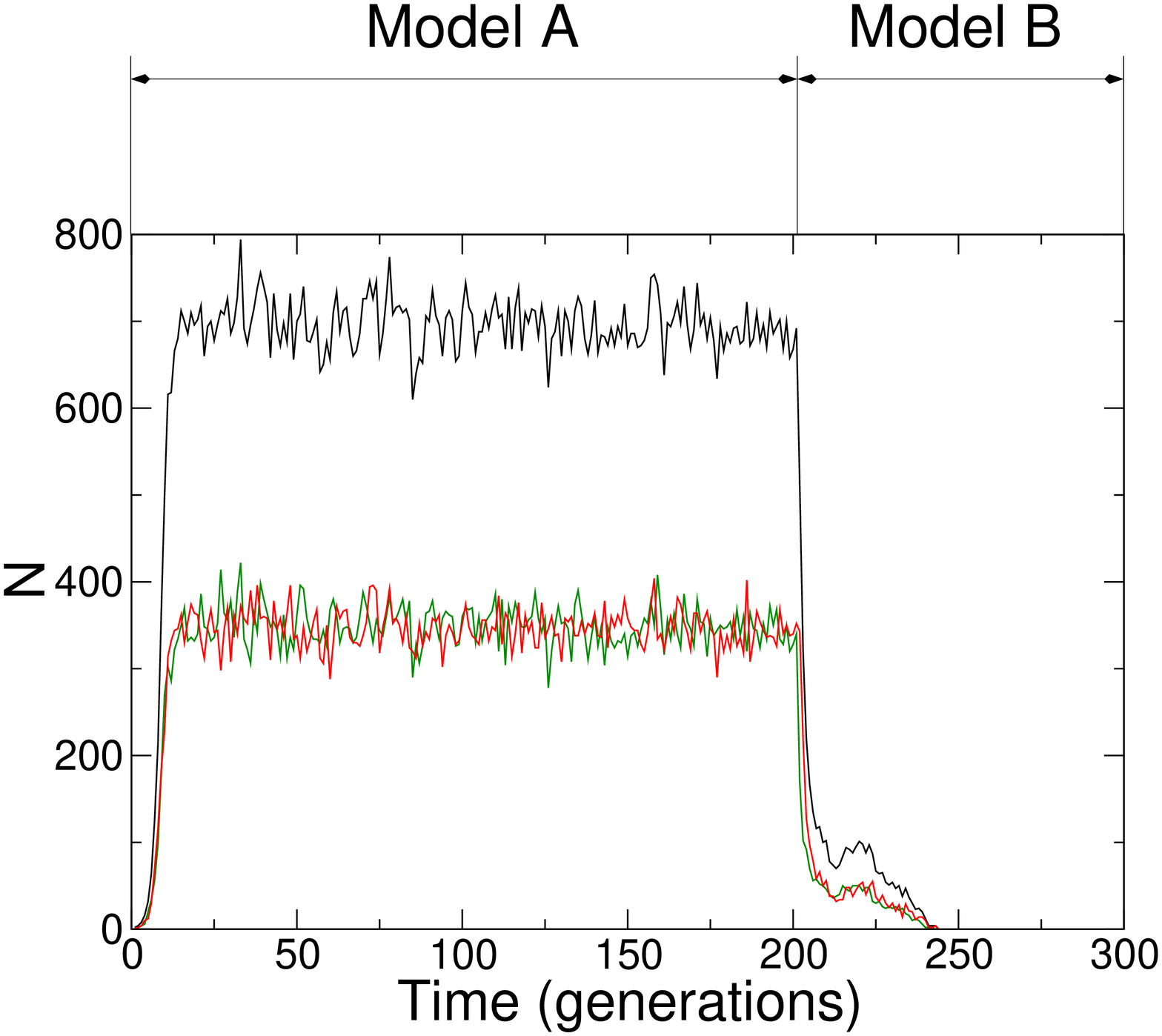} & \includegraphics[scale=0.25]{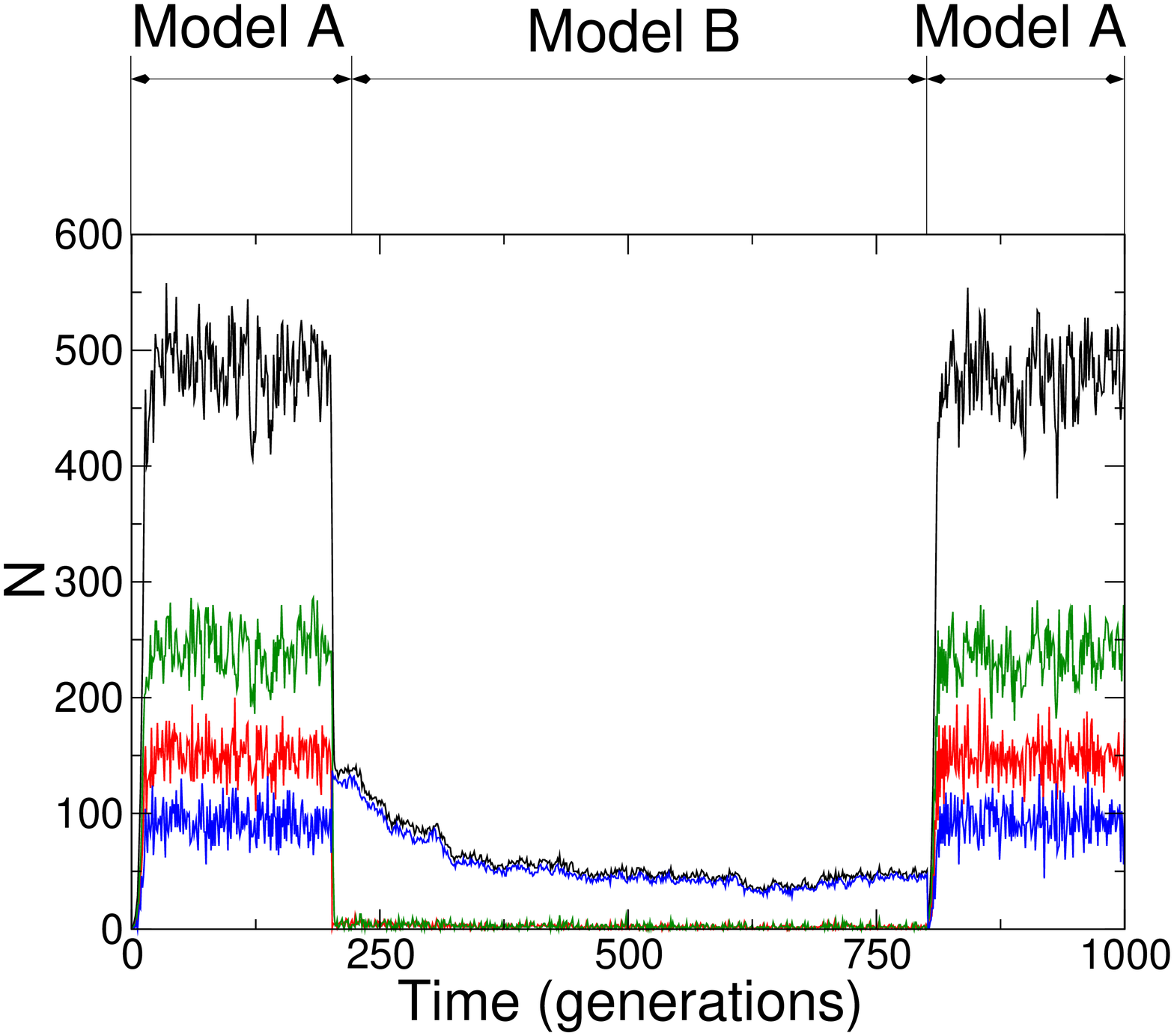}
\end{array}$
\end{center}
\caption{Simulation results showing one realisation of the population dynamics of the reaction to the (transient) application of a therapeutic agent. We can see that in plot (a), which corresponds to a population with no quiescent state, the population upon application of the drug. On the contrary, a population with quiescence is able to survive to the period over which the drug is active, and recovering once the effect of the drug has worn off (plot (b)). Black lines correspond to the total population, red lines to individuals of type (1), green lines to individuals of type (2), and blue lines to quiescent (type (3)) individuals. Parameter values: $r_{12}=r_{21}=0.5$, $r_{13}=r_{31}=0$ for the upper pannel and $r_{13}=r_{31}=0.25$ for the lower pannel, $r_{31}=0.05$, $\mu=0.001$, $\epsilon=10^{-5}$, and $\alpha=1$.\label{fig:3}}
\end{figure*}

\begin{figure}
\includegraphics[scale=0.3]{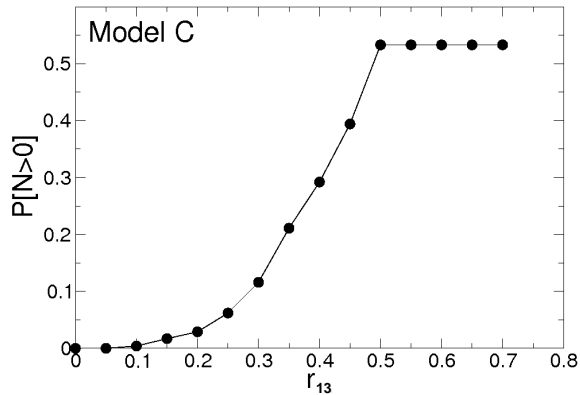}
\caption{Simulation results showing the survival probability, $P[N>0]$, for the population dynamics of the reaction to the (transient) application of a therapeutic agent as a function of the parameter $r_{13}$, i.e. the parameter controlling the flux of population between types 1 and 3. Parameter values: $r_{12}=r_{21}=0.5$, $r_{31}=0.05$, $\mu=0.001$, $\epsilon=10^{-5}$, and $\alpha=1$.\label{fig:4}}
\end{figure}

\begin{figure}
\includegraphics[scale=0.25]{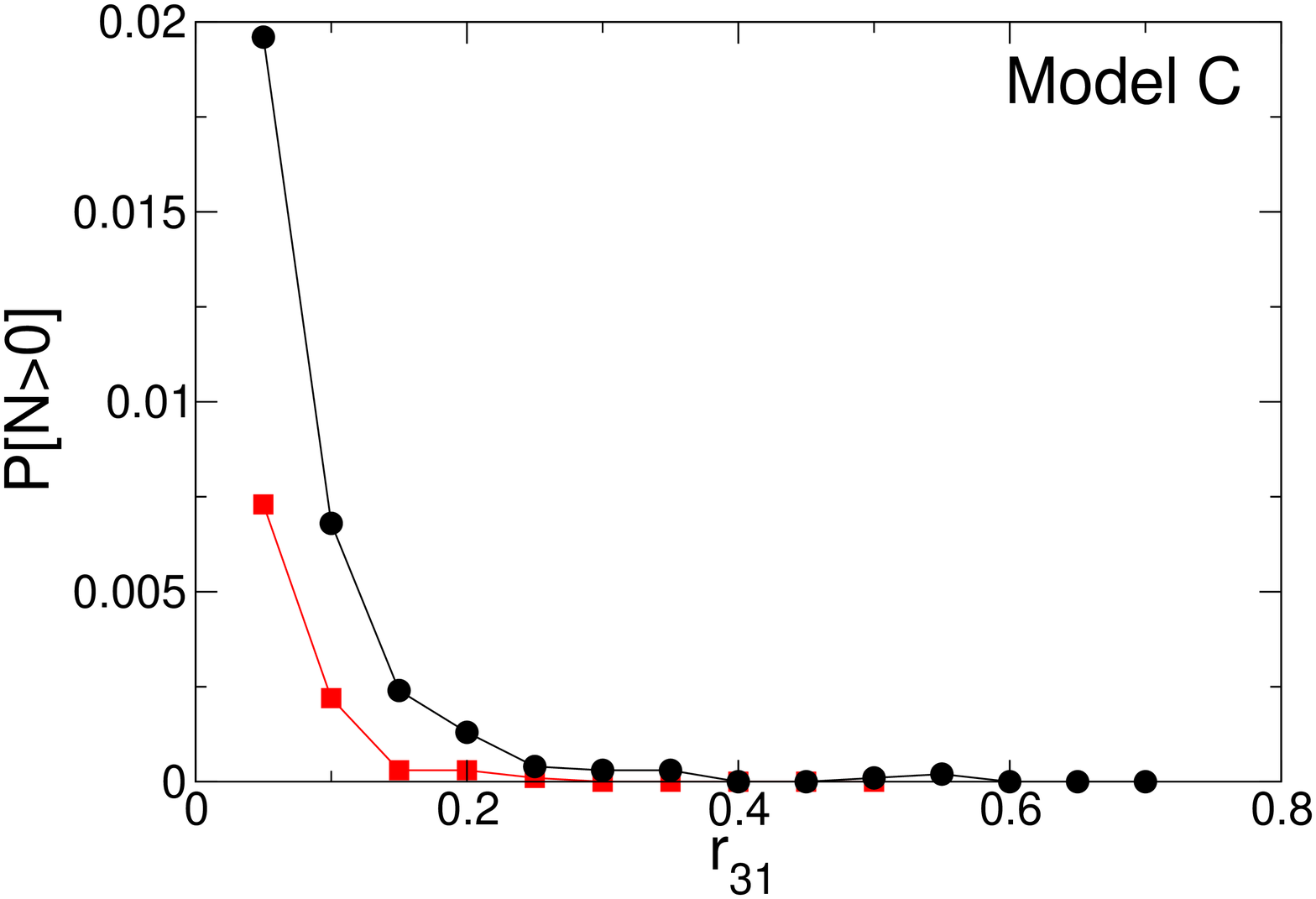}
\caption{Simulation results for model C with $r_{12}=0.51$. Black circles correspond to $r_{13}=0.45$ and red squares to $r_{13}=0.25$. Other parameter values: $\mu=0.001$, $\epsilon=10^{-5}$, and $\alpha=1$.\label{fig:5}}
\end{figure}

To test further this scenario, we perform simulations in which the following situation is considered. First we let a population whose dynamics is given by Model A,  evolve for some time in an environment free of the hostile agent. In this environment, both type (1) and type (2) individuals can thrive. The dynamics in an environment free of any  hostile agent is described in terms of the generating functions shown in Table \ref{table:2}. The introduction of the hostile agent is describe by changing the dynamics of the population to the one described by model B (see the diagram \ref{DiagA} and  Table \ref{table:1}), which is essentially the same as model A (Table \ref{table:2}) but with type (2) individuals doomed to perish. We further assume that the hostile agent is active only for a given period of time, after which its detrimental effect on type (2) organisms  ceases and the dynamics of the population reverts to Model A (as per Table \ref{table:2}). Parameter values are chosen so the dynamics of the agent-free population is super-critical (i.e. the corresponding survival probability is bigger than zero). After the hostile agent has been removed we calculate the probability that the population survives both with and without quiescence. 
This scenario is highly relevant to evolutionary escape from drugs which have a finite life-span. This particular escape problem was considered  Iwasa et al.\cite{iwasa2003}.
 
The corresponding results are shown in Figs. \ref{fig:3} and \ref{fig:4}. Fig. \ref{fig:4} shows the survival probability, $P[N>0]$, as a function of $r_{13}$ and, indeed, we observe that when quiescent individuals are not considered, i.e. $r_{13}=0$, the probability of surviving after the end of the activity period of the hostile agent is null. As $r_{13}$ increases, so does the corresponding survival probability, thus further confirming that quiescence is a feasible mechanism to help biological populations to escape harsh environmental conditions. Moreover, Fig. \ref{fig:3}, which show particular realisations of the process with (panel (b)), and without (panel (a)), quiescence also helps to understand more about the mechanism whereby quiescence allows escape: as can be seen in Fig. \ref{fig:3}, panel (b), during the period of activity of the hostile agent most of the population is actually quiescent with a little proportion of the population in types (1) and (2). This means that quiescence mediates survival by acting as a reservoir or buffer.

\emph{An alternative model.} Whilst the results discussed until here provide compelling evidence in favour of hypothesis that the presence of quiescence can induce escape without the need of an increase in the reproductive fitness of any particular type of organism, some doubt could be cast on our argument so far. One might feel inclined to argue that our model still relays on a multi-type population where one type (type 1) is effectively more  fit  than the others. To adress this possibility, we consider model C in diagram \ref{DiagA}. This version is closely related to the current discussion concerning the response of cancer cells to oxygen starvation. This model is also formulated in terms of a GWMBP and characterised in terms of the generating functions given in Table \ref{table:3}. Its rationale is as follows. Let us assume that a population of cells is divided in two types: those in which the pathways regulating cell division are activated (type 1) and those in which the active pathways are those regulating apoptosis, i.e. cell death, (type 2). Type 1 cells can either proliferate, or stay as they are, or suffering activation of the apoptotic pathways, thus becoming type (2) cells. The flux of population between types (1) and (2) is controlled by the parameter $r_{12}$ . Type 2 cells are those marked for cell death, but the model provides for their staying within the population for a while (not dying immediately), but they cannot proliferate. Type (3) cells are, as in the previous model, quiescent cells. In this context the introduction of a hostile agent (drug, removal of oxygen, etc.) corresponds simply to increasing the value of $r_{12}$. 

In the absence of quiescence, standard arguments reveal that extinction with probability 1 occurs when $r_{12}\geq1/2$. The question is once again whether quiescence can rescue the population from extinction under such conditions. Fig. \ref{fig:5} shows that this is indeed the case, provided that the flux out of the quiescent state back into type (1) does not exceed a critical value. This results are completely analogous to those obtained for the previous model.

\emph{Invasion dynamics.} We know discuss under which conditions a small proportion of individuals, $y$, which can undergo quiescence, modelled by Model A and B depending on whether a drug is present or not, respectively, and with $r_{13}\neq 0$, can take over a population of fully growing individuals, i.e. a population modelled by Model A and B with $r_{13}=0$. In particular, we will analyse the corresponding invasion probability, $P_F(y)$, by direct simulation of the population dynamics and by using the analytical approximation provided by the so-called evolutionary formalism \cite{arnold1994,demetrius1997,demetrius2009,alarcon2009} which allows a more thorough exploration of the behaviour of the invasion probability as a function of the model parameters.

The problem we address here relates to whether a small population of mutants with can take over an incumbent population. Demetrius and coworkers \cite{demetrius1997,demetrius2009} have developed a formalism based on the application of ideas from ergodic theory to evolutionary problems \cite{arnold1994} and the diffusion approximation \cite{feller1951}. This formalism allows us to estimate the fixation probability of a mutant population in the presence of an incumbent species. Here we generalise this formalism to apply it to the problem of whether a genetic inactivation generating new phenotypes can invade the incumbent population.

The starting point of this formalism is the following fundamental equation:

\begin{equation}\label{eq:ef1}
r=H+F,
\end{equation}

\noindent where $r=\log(\Lambda_0)$ is the growth rate (or Malthusian parameter), with $\Lambda_0$ is the dominant eigenvalue of $A$, whereas $H$ and $F$ are the entropy and the proliferative potential, defined as:

\begin{eqnarray}\label{eq:ef2}
\nonumber && H=-\sum_{\imath,\ell}\pi_{\imath}p_{\imath\ell}\ln p_{\imath\ell}\\
&& F=\sum_{\imath,\ell}\pi_{\imath}p_{\imath\ell}\ln A_{\ell\imath}
\end{eqnarray}

\noindent where $p_{\imath\ell}$ is defined as:

\begin{equation}\label{eq:ef3}
p_{\imath\ell}=\frac{A_{\ell\imath}{\cal V}_{\ell}}{\Lambda_0{\cal V}_{\imath}}
\end{equation}

\noindent with ${\cal V}A=\Lambda_0{\cal V}$, and $\pi$ is the stationary distribution associated to ${\cal P}=(p_{ij})$: $\pi_i={\cal V}_i{\cal U}_i$, with ${\cal U}$ given by $A{\cal U}=\Lambda_0{\cal U}$. Eqs. (\ref{eq:ef1})-(\ref{eq:ef3}) are derived from a variational principle \cite{arnold1994}, namely

\begin{eqnarray}\label{eq:ef4}
r=\sup_{\nu\in M}\left(H_{\nu}+\int \gamma d\nu \right)
\end{eqnarray}

\noindent where $\nu$ is a Markov measure $H_{\nu}=-\sum_{j=1}^{n}\nu(A_j)\log \nu(A_j)$ and $\gamma=\log a_{x_0x_1}$, where $A_j,\,j=1,\dots,n$ is a partition of the phase space of the system and $a_{x_0x_1}$ is the transition probability between two states of the system, $x_0$ and $x_1$. The solution to this variational problem produces Eqs. (\ref{eq:ef1})-(\ref{eq:ef3}) \cite{arnold1994,billingsley1965}.

According to Demetrius et al. \cite{demetrius2009} the diffusion approximation yields the following equation for the fixation or invasion probability as a function of the initial concentration of mutants, $y$ is given by:

\begin{equation}\label{eq:dif1}
P_F(y)=\frac{1-\left(1-\frac{\Delta\sigma^2}{\sigma^2_M}y\right)^{\frac{2\langle N\rangle s}{\Delta\sigma^2}+1}}{1-\left(1-\frac{\Delta\sigma^2}{\sigma^2_M}\right)^{\frac{2\langle N\rangle s}{\Delta\sigma^2}+1}}
\end{equation}

\noindent where the total population $N$ is assumed to be constant and $s$ is defined by:

\begin{equation}\label{eq:dif2}
s=\Delta r-\frac{\Delta\sigma^2}{\langle N\rangle}
\end{equation}

\noindent where $\Delta r\equiv r_q-r$, $\Delta\sigma^2=\sigma_q^2-\sigma^2$, with $r_q$ ($r$) is the growth rate of the quiescence (normal) population, and $\sigma_q^2$ ($\sigma^2$) is the variance of the quiescent (normal) population, $\langle N\rangle$ is the average stationary population. 

The demographic parameters (i.e. $r$, $\sigma^2$, $r_q$, and $\sigma_q^2$) can be estimated from the evolutionary formalism (see \cite{demetrius2009,alarcon2009}). The growth rates are given by $r=\log\lambda_0$, where $\lambda_0$ is the dominant eigenvalue of the of the matrix $A=(a_{ij})=(\partial_jG_i(\vec{x})\vert_{\vec{x}=1})$. In \cite{demetrius2009}, it is shown that the parameter $\sigma^2$ can be obtained by slightly perturbing the parameters that determine the dynamics of the system, i.e. the mean-field dynamics being given by $A(\delta)=(a_{ij}^{1+\delta})$, and then doing an expansion for small $\delta$. Accordingly, $\sigma^2$ is given by:

\begin{equation}\label{eq:ef19a}
\sigma^2=-\left.\frac{dH(\delta)}{d\delta}\right\vert_{\delta=0}
\end{equation}

Hence, in the linear approximation with $H(\delta)$ given by $H(\delta)=H+\delta H_{\delta}$, we have that $\sigma^2(\gamma)=-H_{\delta}$, which is given by:

\begin{equation}\label{eq:ef20a}
H_{\delta}=-\sum_{i,j}\pi_i^{(\delta)}p_{ij}(1+\log p_{ij})+\pi_ip_{ij}^{(\delta)}\log p_{ij}
\end{equation}

\noindent where $\pi_i(\delta)=\pi_i+\delta\pi_i^{(\delta)}$ and $p_{ij}(\delta)=p_{ij}+\delta p_{ij}^{(\delta)}$ are the corresponding linear approximations to $\pi$ and $P=(p_{ij})$ when we take $A(\delta)=(a_{ij}^{1+\delta})$ $\delta<<1$. The details of how these quantities are actually calculated is given in Appendix A of \cite{alarcon2009}. 

\begin{figure}
\includegraphics[scale=0.25]{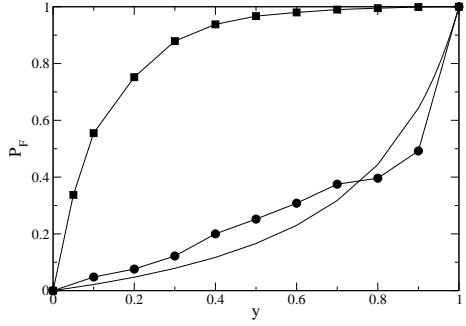}
\caption{Fixation probability for the competition between two populations in the presence of drug. Squares correspond to the competition with a population capable of undergoing quiescence whereas circles correspond to the competition with a population that cannot undergo quiescence/ Solid line corresponds to the analytical expression of $P_F(y)$ for the latter case \label{fixation-probability}}
\end{figure}

In order to show that this formalism can  be used to analyse the invasion of a population possessing a dormant type, we have done simulations where one population, which is sensitive to the drug, competes with two different population: one capable of undergoing quiescence and another one which is unable to enter such state. We compare the numerics with the analytical results obtained from Eq. (\ref{eq:dif1})   in Fig. \ref{fixation-probability}. Both analytical and numerical results show that whereas in the former case the quiescent population takes over the sensitive population almost surely, in the latter case is unlikely that the non-resident population takes over the resident one. It is worth remarking that the good agreement between simulations and analytical result seen in Fig. \ref{fixation-probability}   implies that the evolutionary formalism developed by Demetrius and coworkers  \cite{demetrius1997,demetrius2009}  is able to make the invasion problem considered here analytically tractable. 

\begin{figure}
\includegraphics[scale=0.25]{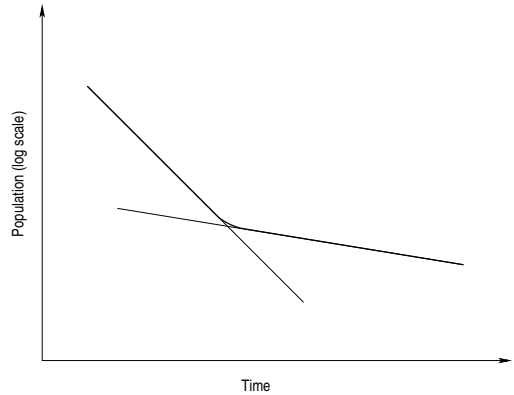}
\caption{This plot illustrates the phenomenon of persistence. After introducing a drug (typically, an antibiotic), there is an initial period of fast decay of the population followed by a cross-over after which the rate of population decays is much slower. In bacterial populations such as \emph{E. Coli}, this is thought to be due to a phenotype switch: Cells switch to a slow growing phenotype that is more resilient to the effect of the grug than the \emph{normal} phenotype. This phenomenon cannot be of genetic origin, as cells regrown from the persistent population upon removal of the drug are sensitive to it \cite{balaban2004}. \label{pers-schem}}
\end{figure}

\emph{Summary and Discussion.} We have shown that quiescence is a feasible mechanism for biological populations to escape hostile environments. This mechanism is expected to be very relevant to the important issue concerning population dynamics of cancer cells and the competition mechanisms between cancer cells and their normal counterparts. In particular our model addresses issues related to the response of cancer cells to oxygen starvation \cite{brown2004,bristow2008}. The mechanism involved in quiescence-dependent escape is in essence  simple: it provides a buffer for the population to be safe from the hostile environment. I.e. in cases where the population would go extinct due to a fluctuation in the birth and death events the reservoir in the quiescence buffer population can bring the population back from the brink of extinction.
The efficiency of this mechanism for escape is sensitive to the entry and exit rates to and from the quiescent state, respectively.

The main difference between our mechanism for escape and the one proposed in \cite{iwasa2003,iwasa2004} is the following. Iwasa and co-workers propose a mechanism based on random search of the state of types of individuals for one that is fitter than the others in the presence of a given selection prossure (drug, etc.).  In contrast we have demonstrated here that the overall survival probability of a population may increase by introducing a dormant type. Since this type is unable to reproduce its reproductive fitness is zero, nevertheless the existence of this drug resistant stage is able to improve the fitness of the entire population in as much as the population obtains a higher probability for survival. 

The mechanism proposed here shares a number of features in common with the phenomenon of bacterial persistence \cite{balaban2004,levin2006,gardner2007,lewis2007}. Persistence is a form of resistance to antibiotics exhibited by bacterial colonies where resistance is not acquired by a gene mutation which allows the bacteria to grow exponentially fast in the presence of a particular drug. Instead, as schematically shown in Fig. \ref{pers-schem}, killing of bacteria goes through a fast phase where the population decreases exponentially until the killing rate slows down leaving behind a remnant of cells which differ from the sensitive ones in that they are in a dormant state but are otherwise genetically identical to their sensitive counterparts. In fact, when the drug is removed, a colony of ``normal'' bacteria is regrown from the persistent cells. The mechanism for persistence appears to involve a phenotypic switch \cite{balaban2004} where cells switch from a rapidly growing, but drug-sensitive phenotype to a dormant but drug-resistant phenotype, although there has recently been the suggestion that persistence might be a social trait \cite{gardner2007}. 

According to \cite{balaban2004}, there exist two types of persisters. Type I persisters, according to their terminology, are only produced during the exponential growth phase and therefore their numbers are fixed at the time of innoculation and determined by the size of the innocolum. Type II persisters, on the contrary, divide and grow continuously, but an order of magnitude slower than their non-persistent counterparts, and their numbers are determined by the total population numbers. The quiescence mechanism put forward here is therefore distinct from type I persisters, but very similar to the behaviour exhibited by type II persisters, which means that the analysis methods, in particular, the evolutionary formalism used to study the competition between sub-populations exhibiting different behaviour, can be extended to study this type of bacterial persistence.

Another issue in which our model differs from previous work on bacterial persistence is the following. We are consider different populations adopting different strategies, namely, a wild-type population which thrives in the absence of drug composed of two subpopulations: one that is capable of undergoing quiescence and another one that is not. By adopting this scenario we can study the evolutionary dynamics of these populations and analyse which strategy is evolutionary stable and which one is susceptibel to be invaded. In that respect, our analysis goes beyond, for example, the population models proposed by \cite{balaban2004}.

Within the context of the problem of how the hypoxic sub-population affects the dynamics of the whole tumour, our model sheds further light on this issue. It is commonly thought that hypoxia increases the probability of survival of the tumour by providing a selective pressure that favours the evolution and survival of more aggresive phenotypes \cite{bristow2008}. Our model shows that quiescence by itself is enough to increase the survival probability of the population without further increase in the phenotypic variety of the population. Obviously, the residual population provides a springboard for these evolutionary processes to ensue. 

TA and HJJ gratefully acknowledge the EPSRC for funding under grant EP/D051223.   
 
\appendix

\section{Simulations}

In this appendix we briefly describe the method we have used to produce our simulation results. Our simulations start with one single individual of type 1. Its offspring is the determined by the probabilities of producing descendants as prescribed by the generating functions given in Tables \ref{table:2}, \ref{table:1}, and \ref{table:3}, corresponding to Models A, B, and C, respectively. In general, the coefficients of the Taylor expansion of $G_i(\vec{x})=\sum_{l,n,m=0} P_{i;lnm}x_1^lx_2^nx_3^m$ are the probabilities per generation per indvidual (of type $i$) that an type $i$-individual produces $l$ descendants of type $1$, $n$ descendants of type $2$, and $m$ descendants of type $3$. In subsequent generations, we go over all the individuals in existance in the last generation and the numbers and types of their descendants within the next generation are calculated in the same way. Simulations are run over 1000 realisations of 1000 generations each. The survival probability $P[N>0]$ is calculated as the ratio between the number of simulations such that $N(T=1000)>0$ and the total number of realisations.

\begin{table*}
\caption{{\bf Model A}. Generating functions of the probabilities of the number of offspring for individuals belonging to our population previous to the introduction of a hostile agent. In this situation, both type 1 and type 2 indviduals are well adapated and can proliferate subject to the restrictions imposed by the carrying capacity. Type 3 correspond to quiescent individuals.  The new parameters introduced here are $r_{21}$, i.e. the mutation probability of type 2 into type 1, the probability of a type 2 individual to enter quiescence, $r_{23}$, and the probability of a quiescent individual to revert to type 2. The rest of the parameters have the same physical meaning as in Table \ref{table:1} \label{table:2}}
\begin{ruledtabular}
\begin{tabular}{l|c}
Population 1 & $G_1(\vec{x})=1-e^{-\mu N}+e^{-\mu N}r_{12}x_2^2+e^{-\mu N}r_{13}x_3^2+e^{-\mu N}(1-r_{12}-r_{13})x_1^2$ \\ \hline
Population 2 & $G_2(\vec{x})=1-e^{-\mu N}+e^{-\mu N}r_{21}x_2^2+e^{-\mu N}r_{23}x_3^2+e^{-\mu N}(1-r_{21}-r_{13})x_1^2$ \\ \hline
Population 3 & $G_3(\vec{x})=\epsilon + r_{31}e^{-\mu N}x_1 + r_{32}e^{-\mu N}x_2+ (1-\epsilon - (r_{31}+r_{32})e^{-\mu N})x_3$ \\
\end{tabular}
\end{ruledtabular}
\end{table*}

\begin{table*}
\caption{{\bf Model B.} Generating functions of the probabilities of the number of offspring for individuals of type 1, i.e. well adapated to harsh conditions, $G_1(\vec{x})$, type 2, i.e. those for which the environment is lethal, $G_2(\vec{x})$, and type 3, i.e. quiescent individuals, $G_3(\vec{x})$.  The parameter $\mu$ is the carrying capacity, which accounts for the limitation in resources, $\alpha$ is a measure of the remanent resilience of inviduals of type 2 to the environment, and $\epsilon$ is the death rate (probability of death per individual per generation) of the quiescent cells. The quantity $r_{1j}$ with $j=1,2$ can be interpreted as the muation rate of individuals of type 1 into individuals of type $j$. The quantity $r_{31}$ is the probability per quiescent individual per generation of a type 3 individual to revert to ype 1. \label{table:1}}
\begin{ruledtabular}
\begin{tabular}{l|c}
Population 1 & $G_1(\vec{x})=1-e^{-\mu N}+e^{-\mu N}r_{12}x_2^2+e^{-\mu N}r_{13}x_3^2+e^{-\mu N}(1-r_{12}-r_{13})x_1^2$ \\ \hline
Population 2 & $G_2(\vec{x})=\alpha  + (1-\alpha )x_2$ \\ \hline
Population 3 & $G_3(\vec{x})=\epsilon + r_{31}e^{-\mu N}x_1 + (1-\epsilon - r_{31}e^{-\mu N})x_3$ \\
\end{tabular}
\end{ruledtabular}
\end{table*} 

\begin{table*}
\caption{Generating functions corresponding to model C (see main text for details). Type 1 and type 2 correspond to the proliferative and apoptotic phenotypes, respectively. Type 3 corresponds to the quiescence population.  \label{table:3}}
\begin{ruledtabular}
\begin{tabular}{l|c}
Population 1 & $G_1(\vec{s})=(1-e^{-\mu N})x+e^{-\mu N}r_{12}x_2^2+e^{-\mu N}r_{13}x_3^2+e^{-\mu N}(1-r_{12}-r_{13})x_1^2$ \\ \hline
Population 2 & $G_2(\vec{s})=\alpha  + (1-\alpha)x_2$ \\ \hline
Population 3 & $G_3(\vec{s})=\epsilon + r_{31}e^{-\mu N}x_1+ (1-\epsilon - r_{31}e^{-\mu N})x_3$ \\
\end{tabular}
\end{ruledtabular}
\end{table*}

\end{document}